# An exploratory study of Google Scholar

Philipp Mayr[1] and Anne-Kathrin Walter
*GESIS / Social Science Information Centre (IZ), Bonn, Germany*

**Abstract**
Purpose – This paper[2] discusses the new scientific search service Google Scholar (GS). This search engine, intended for searching exclusively scholarly documents, will be described with its most important functionality and then empirically tested. The focus is on an exploratory study which investigates the coverage of scientific serials in GS.

Design/methodology/approach – The study is based on queries against different journal lists: international scientific journals from Thomson Scientific (SCI, SSCI, AH), Open Access journals from the DOAJ list and journals of the German social sciences literature database SOLIS as well as the analysis of result data from GS. All data gathering took place in August 2006.

Findings – The study shows deficiencies in the coverage and up-to-dateness of the GS index. Furthermore, the study points up which web servers are the most important data providers for this search service and which information sources are highly represented. We can show that there is a relatively large gap in Google Scholar's coverage of German literature as well as weaknesses in the accessibility of Open Access content. Major commercial academic publishers are currently the main data providers.

Research limitations/implications – Five different journal lists were analyzed, including approximately 9,500 single titles. The lists are from different fields and of various sizes. This limits comparability. There were also some problems matching the journal titles of the original lists to the journal title data provided by Google Scholar. We were only able to analyze the top 100 Google Scholar hits per journal.

Practical implications – We conclude that Google Scholar has some interesting pros (such as citation analysis and free materials) but the service can not be seen as a substitute for the use of special abstracting and indexing databases and library catalogues due to various weaknesses (such as transparency, coverage and up-to-dateness).

Originality/value – We do not know of any other study using such a brute force approach and such a large empirical basis. Our study can be considered as using brute force in the sense that we gathered lots of data from Google, then analyzed the data in a macroscopic way.

Paper type – Research paper

---

[1] Contact Philipp Mayr at: mayr@iz-soz.de
[2] A revised and updated version of the presentation "Google Scholar – How deep does this search engine dig?" delivered at the IuK-Conference, 2005 in Bonn. We wish to thank our colleagues Max Stempfhuber, Peter Mutschke, Jürgen Krause and Vivien Petras for their assistance in preparing this paper. This paper is part of the BMBF funded project "Competence Center Modeling and Treatment of Semantic Heterogeneity"
grant no. 523-40001-01C5953. See http://www.gesis.org/en/research/information_technology/komohe.htm

# Introduction

As is now customary for new Google offerings, the launch of Google Scholar (http://scholar.google.com/) generated a great deal of media attention shortly after its debut in November 2004. Its close relation to the highly discussed topics of open access and invisible web (Lewandowski and Mayr, 2006) ensured that many lines were devoted to this service in both the general media (Markoff, 2004; Terdiman, 2004) and among scientific publishers and scientific societies (Banks, 2004; Butler, 2004; Payne, 2004; Sullivan, 2004; Jacsó, 2004; Giles, 2005). While the initial euphoria over this new service from Google has since quieted down, the service is currently being utilized by academic search engines to integrate results that are available free of charge.

Google Scholar stands out not just for the technology employed but for the efforts made to restrict searches to scientific information. As stated on the Google Scholar webpage:

*"Google Scholar enables you to search specifically for scholarly literature, including peer-reviewed papers, theses, books, preprints, abstracts and technical reports from all broad areas of research. Use Google Scholar to find articles from a wide variety of academic publishers, professional societies, preprint repositories and universities, as well as scholarly articles available across the web."*
(Google 2005, see http://scholar.google.com/scholar/about.html)

Above all, it appears that Google is attempting to automatically index the totality of the realm of scientifically relevant documents with this new search service Google Scholar. As Google does not make any information available with regard to coverage or how current the content it offers is, this study has been undertaken with the goal of empirically exploring the depth of search in the scientific web. We have measured the coverage of the service by testing different journal lists. The types of results and which web servers are represented in the result are also analyzed.

The paper first describes the background, functions and unique features of Google Scholar. A brief literature review will bring together the current research results. Results of the second Google Scholar study from August 2006 will be presented in the second part. An initial analysis of journals in Google Scholar was conducted by the authors in the period April/May 2005 (Mayr and Walter, 2006). The results of this study were compared with certain parts of the current analysis in August 2006. This is followed by a summary of our observations on this new service.

# Google Scholar

The pilot project CrossRef Search (http://www.crossref.org/crossrefsearch.html) can be seen as a test and predecessor of Google Scholar. For CrossRef Search Google indexed full-text databases of a large number of academic publishers such as Blackwell, Nature Publishing Group, Springer, etc., and academic/professional societies such as the Association for Computing Machinery, the Institute of Electrical and Electronics Engineers, the Institute of Physics, etc., displaying the results via a typical Google interface. The CrossRef Search interface continues to be provided by various CrossRef partners (e.g. at Nature Publishing Group).

Similar in approach, but broader and less specific in scope than Google Scholar, the scientific search engine Scirus (http://www.scirus.com) searches, according to information they provide, approximately 300 million science-specific web pages. In addition to scientific documents from Elsevier (ScienceDirect server, see http://www.sciencedirect.com/) freely accessible documents, many from public web servers at academic institutions, are provided. Among these are, for example, documents placed by students that do not fulfil scientific criteria such as peer review which often lead to their exclusion in searches. In our experience there is more than a negligible fraction of records from non-academic web spaces in the Scirus index. Scirus' coverage of purely scientific sources in addition to Elsevier's ScienceDirect full-text collection is low by comparison (compare the selection of hosts in the Scirus advanced search interface, http://scirus.com/srsapp/advanced/). What Scirus declares as the 'rest of the scientific web' is too general, non-specifically filtered and makes up the majority of hits in any query.

As seen in the pilot project CrossRef Search, the chosen Google Scholar approach is to work in cooperation with academic publishers. What is significant about the Google Scholar approach?

First and foremost, what stands out is that Google Scholar, as previously mentioned, delivers results restricted to exclusively scientific documents and this constraint has yet to be consistently implemented by any other search engine. Google Scholar is a freely available service with a familiar interface similar to Google Web Search. Much of the content indexed by Google Scholar is stored on publishers' servers where full-text documents can be downloaded for a fee, but at least the abstracts of the documents found will be displayed at no cost. The Google approach does, however, provide documents from the open access and self-archiving areas (compare Swan and Brown, 2005).

In addition to the full-text access users might also be interested in the analysis implemented by Google and the document ranking based on this analysis. The relevance ranking is based on various criteria (see citation below). According to this the citation value of a document is only one factor contributing to its ranking. Google builds a citation index out of the full-text index as an add-on to its service. On top of the statistical best match ranking of full-texts, this add-on implementation can be valuable for re-ranking documents or for analysis and evaluation purposes of certain document sets. Automatic reference extraction and analysis, also known as Autonomous Citation Indexing (ACI), can be particularly helpful for the user in information retrieval and delivery. This process ensures that often cited scientific works will be ranked more highly in the results list thereby making them more visible to the user. Additionally the user can track all citing works extracted by ACI which need not necessarily be included in the full-text index or contain the original user search term. The automatic ACI process necessitates that references in the documents analyzed be available, which is, per se, for granted if full-texts are analyzed. This procedure also enables Google Scholar to present additional references not found on the indexed web servers.

Figure 1 is a graphic representation of the Google Scholar approach including the value added service ACI. The three different citing styles for the same reference seen in figure 1 are taken from Lawrence, Giles & Bollacker (1999) and are intended to illustrate the difficulties in dealing with automatic normalization of references. The original system CiteSeer (http://citeseer.ist.psu.edu/) as well as Google Scholar have up to now implemented only heuristics for the application of ACI that also produce some errors in the citation values (see also Jacsó, 2005c; 2006a; 2006b).

# take in Figure 1:

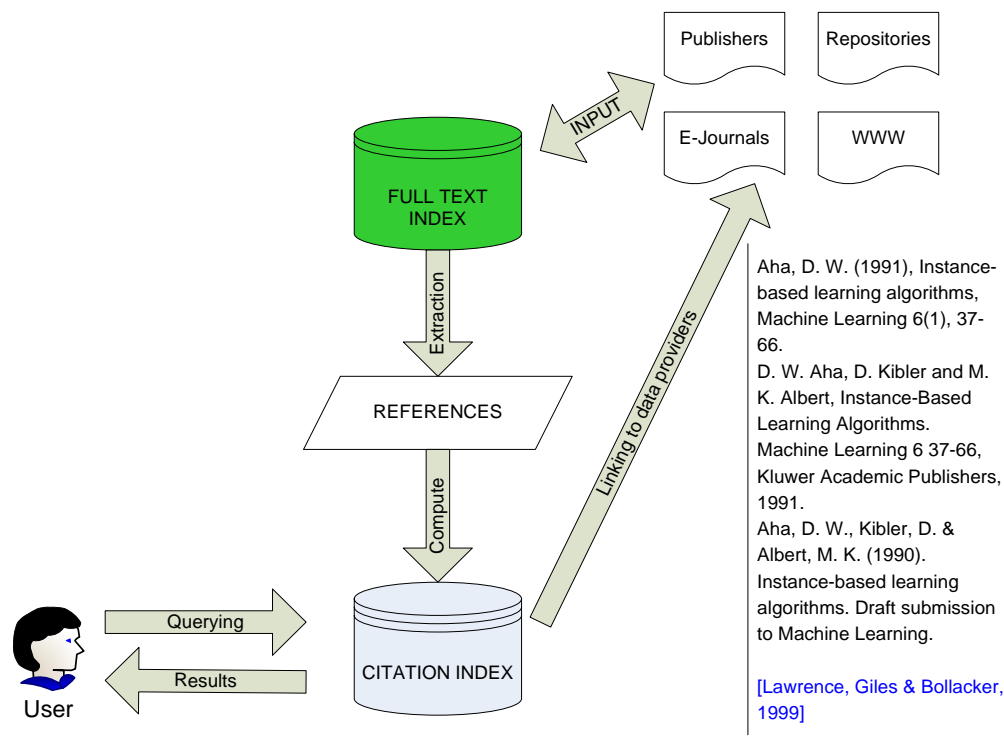

Figure 1: Google Scholar Approach

Google Scholar is also noteworthy for the fact that it is conceived of as an interdisciplinary search engine. In contrast to specialty search engines like the CiteSeer system which indexes freely available computer science literature or RePEc for economic papers, the Google Scholar approach can be conceived of as a comprehensive science search engine.

The following is a short description of the most important features of Google Scholar:

- *Advanced search:* The advanced search offers, in addition to searching the title of the article, the opportunity to search for an author name, journal title and year of publication of an article or book (see Jacsó, 2005a; 2005b for details on the limitations). These attributes represent only a minimal set of search criteria compared to specifically scientific search interfaces and the reliable extraction of this data from un- or only partially-structured documents poses a serious problem for an automatic system. The advanced search has recently begun to offer access by subject to different disciplines.
- *Full text access:* In contrast to the classical abstracting and indexing databases, which search in bibliographic metadata, including abstract and keywords, Google Scholar searches based on a full-text index. This means that the user can – with minor limitations (Price, 2004) and all the advantages and disadvantages of this kind of search – directly search and access the full text of documents.
- Relevance ranking: Google states (2004): "Just as with Google Web Search, Google Scholar orders your search results by how relevant they are to your query, so the most useful references should appear at the top of the page. This relevance ranking takes into account the full text of each article as well as the article's author, the publication in which the article appeared and how often it has been cited in scholarly literature. Google Scholar also automatically analyzes and extracts citations and presents them as separate results, even if the documents they refer to are not online. This means your search results may include citations of older works and seminal articles that appear only in books or other offline publications." (Google 2004, see

http://web.archive.org/web/20041130084532/scholar.google.com/scholar/about.html) The relevance statement offered by Google in 2004 has since been shortened to the following: "Google Scholar aims to sort articles the way researchers do, weighing the full text of each article, the author, the publication in which the article appears, and how often the piece has been cited in other scholarly literature. The most relevant results will always appear on the first page." (Google 2007, see http://scholar.google.de/intl/en/scholar/about.html)

- *Web Search:* The link to the Google main index is useful especially when the documents are not directly available from the Google Scholar result list and the query is expanded to the whole (Google) web.
- *Institutional Access:* The pilot project Institutional Access mainly offers additional value for institutional users such as students or scientific staff as Google uses open linking/ link resolver such as SFX to link directly to local library holdings.
- *Additional Features:* Google Scholar offers additional features like *Library Search* which links the query to OCLC WorldCat (http://www.oclc.org/worldcat/) thereby providing hits from local libraries. Alternative places of a document on the web will also be presented (see Fig. 2 *versions*).

Figure 2 shows a typical Google Scholar results list. The individual components of a hit will be discussed in more detail later. Figure 2 illustrates that the availability of a hit can differ. The two different items depicted in the figure (labeled as book or citation) are not accessible via hyperlink as they are extracted only from indexed documents.

# take in Figure 2

Figure 2: Google Scholar results list. The search was for titles containing the phrase "digital library."

## How deep does Google Scholar dig?

Much criticism has already been leveled at the lack of information about the actual size and coverage of Google Scholar (Jacsó, 2004; 2005a; 2005b; Mayr and Walter, 2006). Remaining questions as to how often the search engine index is truly updated can not be answered from publicly accessible research sources.

We would like to preface our journal title study of Google Scholar by giving a brief literature review of related studies published since the launch of Scholar. In our view there are at least two types of literature attempting to challenge Google Scholar in an academic way. There are papers analyzing the functionality, coverage and up-to-dateness of the Scholar service and there are studies using Scholar as an instrument and alternative tool for citation analysis.

Peter Jacsó began early on with his reviews of Scholar. In his critical commentaries (2004, 2005a, 2005b, 2005c) he pointed out that important features of academic search services like accurate searching of journal names (including name abbreviations), Boolean logic or publication years can be quite annoying and contain lots of mistakes in Scholar. The same problems arise in trying to count citations or hits (2006a, 2006b).

*"Those who need a comprehensive set of papers that includes the most respected (and hence most-cited) articles, books and conference papers are advised to treat the hits — and citedness scores — in Google Scholar with much reservation."* (Jacsó, 2005b)

His observations lead him to conclude that Scholar could be a useful service if its implementation would be more careful and elaborated, but in its current beta status Scholar is not sufficient for scholarly research.

Beside arguments of functionality and accuracy, in our eyes there are the increasingly critical points of size, coverage, completeness and up-to-dateness to be noted when using Scholar as a search tool. Google fails here, because it gives too little information about its sources. Some other researchers and professional searchers analyzing size, coverage, etc. have also registered their concerns about this policy (Noruzi, 2005; Bauer and Bakkalbasi, 2005; Mayr and Walter, 2006).

*"However, it is important for all researchers to note that until Google Scholar gives a full account of what material it is indexing and how often that index is updated, it cannot be considered a true scholarly resource in the sense that Web of Science and Scopus are. An understanding of the material being covered is central to the validity of any search of scholarly material."* (Bauer and Bakkalbasi, 2005)

It can be said that Google Scholar covers only a part of the indexed document collections. The extent of this difference is often great (see Jacsó, 2005c), but it is difficult to explain it in a statistical correct way (compare Mayr and Tosques, 2005 for analyses with the Google APIs web service). We assume that Google Scholar has started by only indexing a part of holdings. Preliminary and non-representative results of these experimental studies – including author, journal or topical searches – underscore the beta status of the Google Scholar service leading to the conclusion that presently the index is irregularly updated and completeness and up-to-dateness varies greatly between different collections.

Google Scholar is also drawing attention from literature coming from the fields of bibliometrics and informetrics. Researchers from this field compare the new Google Scholar service with the established citation indices Web of Science (WoS) and Scopus (Bauer and Bakkalbasi, 2005; Belew, 2005; Noruzi, 2005; Kousha and Thelwall, to appear) or other citation databases (e.g. CiteSeer, see Bar-Ilan, 2006). Most of these studies are basing on small samples and applying different methodologies. Bauer and Bakkalbasi stated that 'Google Scholar provided statistically significant higher citation counts than either Web of Science or Scopus', but this result is based on the analysis of only one journal and two different journal volumes. They also say that older material from the analyzed journal is covered better by WoS. Belew (2005) applauds the 'first independent confirmation of impact data' but also identifies significant variations in the counts between the ISI/WoS and the Google citation database. Belew and Bauer and Bakkalbasi also mentioned that Google Scholar could possibly cover the Open Access/self archiving web publishing fraction better than the traditional citation activity WoS. Noruzi (2005) compared citations counts for highly cited papers in the webometrics field. He found a certain overlap between Scholar and WoS and a good ratio of additional papers for Google Scholar. Kousha and Thelwall (to appear) compared traditional and web-based citation patterns of Open Access articles in multiple disciplines. They found 'significant correlations and overlaps between ISI/WoS citations and both Google Scholar and Google Web/URL citations' in all disciplines studied. Correlation between ISI/WoS citations and Google Scholar citation are stronger than ISI/WoS correlated with Google Web citations. Kousha and Thelwall concluded from their interesting study that it could be said that Google Scholar had a 'widely applicable value in citation counting,' but that Scholar's limitations must also be noted.

Our study was carried out as an alternative attempt to create a more accurate picture of Google Scholar' current situation. Compared with the former studies, it utilizes a brute force approach to give a more macroscopic view on the content indexed by Scholar. Our study uses brute force in the sense that we gathered a lot of data from Google, and analyzed the data in a macroscopic fashion. The following study addresses the question: How deep does Google Scholar dig? The study should make it possible to answer these research questions:

- How complete is Google Scholar's coverage of different scientific journals on a general level? By querying multiple journal lists the study tests whether Google Scholar has indexed the journals and can display the articles. The journal lists come from widely varying subject areas: international peer-reviewed journals from the Web of Science (http://scientific.thomson.com/products/wos/) (particularly Science, Technology & Medicine), Open Access and social sciences, and enable conclusions to be drawn about the thematic focus of the current Google Scholar offering. Is Scholar touching the academic invisible web (compare Lewandowski and Mayr, 2006)?
- Which document types does Google Scholar deliver? Are theses results sufficient for professional searchers and academic researching? The analyzed data gives indications about the composition and utility of the results delivered by Scholar: full-text, link and citation.
- From which providers does Google Scholar take the bulk of the documents retrieved? The study should show who the most prominent providers of data for this new service are, and which sources for scientific information are actually underrepresented in the index. The distribution of the web servers and providers is significant as it is an indicator of whether Google Scholar delivers more pay per document or freely accessible documents.

## Methodology

In August of 2006 five different journal lists were queried and the results returned were analyzed. In most scientific disciplines journals are the most important forum for scientific discussion; they can be readily processed and a relatively small amount of journals yields a representative and evaluable amount of results.

Since not all existing journals could be queried, a selection was made from these readily available journal lists.

- Journal lists from *Thomson Scientific* (ISI, see http://scientific.thomson.com/mjl/).
    - Arts & Humanities Citation Index (AH = 1,149 Titles) contains journals from the Humanities
    - Social Science Citation Index (SSCI = 1,917 Titles) contains international social science journals[3]
    - Science Citation Index (SCI = 3,780 Titles) contains journals from Science/Technology and Medicine
- Open Access journals from the *Directory of Open Access Journals* (DOAJ, see http://www.doaj.org/). At the time of the study this list encompassed a total of 2,346 international Open Access Journals from all scientific fields.
- *Journals from the SOLIS database* (IZ, Sozialwissenschaftliches Literaturinformationssystem, see http://www.gesis.org/Information/Zeitschriften/index.htm. This list encompasses a total of 317 mainly German language journals from various sociological disciplines and related areas.

The five journal lists cover very different areas and cannot be directly compared in terms of content, range, and size. More insight should be gained regarding which scientific disciplines, in what form and to what depth can be reached by Google Scholar. It should be noted that the five journal lists analyzed reflect only a small number of regularly appearing journals. The Electronic Journals Library (http://www.bibliothek.uni-regensburg.de/ezeit/) in Regensburg, Germany for example, covers more than 22,800 periodical titles, of which more than 2,650 are purely online journals. *Harnad et al.* (2004) arrive at a figure of approximately 24,000 peer-reviewed journals. Other estimates set the figure at about 100,000 periodically appearing publications (Ewert and Umstätter, 1997).

The study is divided into the following steps:

- *Step 1:* Querying the journal titles: Titles from all journal lists were queried to determine the coverage of Google Scholar. The aforementioned lists were queried in August, 2006. Advanced search offers the field "Return articles published in …."

- *Step 2:* Downloading of Google Scholar result pages: A maximum of 100 records were downloaded for every journal title to be processed.

- *Step 3:* Data extraction from the results list: The data studied is based on the individual records of the results pages. To clearly illustrate the approach, the typical structure of a Google Scholar hit is described in the following paragraph below.

- *Step 4:* Analysis and aggregation of the extracted data. The extracted data was aggregated using simple counts. We first counted each journal whose title could either be clearly identified or not. The results which could be matched were ordered according to the four different types of documents and counted (see Fig. 3). For each result matched to a journal, all domains were extracted and the frequency of the individual web servers per journal list was computed (see Table 3).

---

[3] The definition of social sciences in SSCI is known to be rather broad and even contains, for example, information science journals.

## Composition of Google Scholar Records

Figure 3 shows the components of a typical Google Scholar record:

- Title and document type of the record
- Domain of the web server
- Citation count of the document
- Journal title

# Take in Figure 3

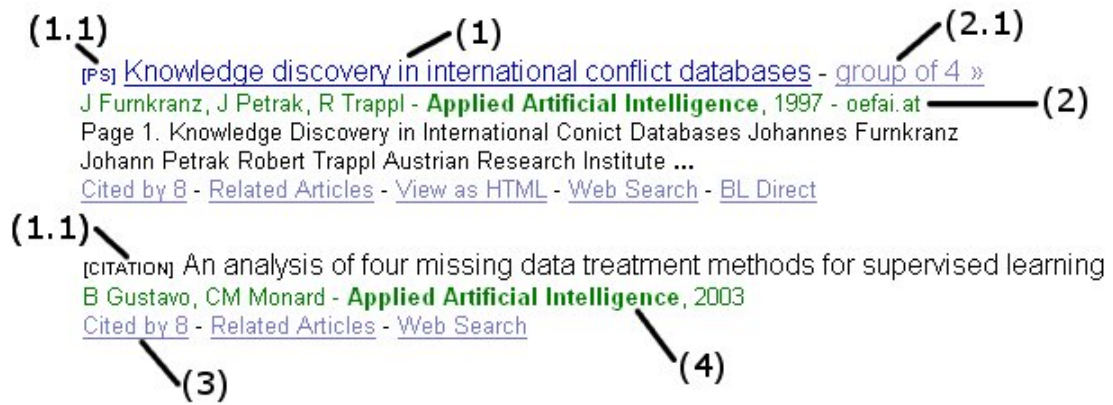

Figure 3: Two typical records of a Google Scholar result. Search was for the journal *Applied Artificial Intelligence*.

**Title and document type of the record (1)**

In addition to the relevance of a reference users are also interested in the availability of documents. The best case scenario is when users are directly linked to the full text; less favorable is when only a citation is displayed with the opportunity to query further via Google Web Search. The first line determines the type of the record. Certain types of documents are marked by brackets in front of the actual title to indicate their type.

- Direct link to full text in Postscript- or PDF-Format: Indicates a full-text record in Postscript or PDF-Format; "PS" or "PDF," respectively, appearing as prefix in brackets (1.1 in Fig. 3). This is not always the case for PDF-files so the suffix of the link must also be taken into consideration.
- "Normal" reference: Most of the records are links, leading first to a bibliographic reference which, according to Google Scholar, should contain at least one abstract.
- Citations: Many journal articles are offered by Google Scholar only as a citation. These results are denoted by the attached prefix "CITATION" (1.1 in Fig. 3) and are not backed up by a link.
- Books: Google Scholar also delivers books as results, denoted by "BOOK." As this study is only concerned with references found in journals these will not be considered.

**Domains (2)**

If the record is a link, the main web server is denoted (see 2 in Fig. 3). If there are multiple sources, these can be reached by clicking the link "group of xy" (see (2.1) in Fig. 3). These links were not included in the analysis; we only analyzed the main link for each linked record.

**Citation count (3)**

Document ranking by Google Scholar is partially based on article citation counts. These are displayed (see (3), or "Cited by xy" in Fig. 3) but were not evaluated for this study.

**Journal title (4)**

Google Scholar supports phrase search in limited fashion so journals will be searched and displayed which do not necessarily contain the search term as a phrase. For this reason every record was individually checked and only counted as a hit when the exact title (see (4) in Fig. 3) was found.

## Results

### 1. Identification of journals

First, we checked how many journal titles from the lists could be identified by Google Scholar. Journals were only classed as "Titles found" when they were clearly identifiable on the returned data. All titles not clearly identifiable were labeled as "Titles not found."

Table 1: Identification of journal titles in Google Scholar data

| List | Titles | Titles found (in %) |
|---|---|---|
| AH | 1,149 | 925 (80.50) |
| DOAJ | 2,346 | 1,593 (67.90) |
| IZ | 317 | 222 (70.03) |
| SCI | 3,780 | 3,244 (85.82) |
| SSCI | 1,917 | 1,689 (88.11) |

Table 1 shows that the majority of requested journal titles from the five lists (AH, DOAJ, IZ, SCI, SSCI) can be identified in the data delivered from Google Scholar (see Titles found column; average is round 78.5%), and that articles in the journals could actually be found. The exact number of the individual articles of a journal could not be determined because our analysis included only 100 hits for each journal. From the 317 journals on the IZ journal list (SOLIS) 222 titles (about 70% of the list) can be clearly identified (see "Titles found"). The remaining 30% of the list can not be clearly identified, or produce no hits. There was, interestingly, a relatively high number of journal titles found for all lists. Yet, surprisingly, only 67.9% of the freely accessible, open access journals can be definitively identified (see DOAJ list). The values of the DOAJ lists have fallen by about 10% when compared with our previous study in April/May 2005 (Mayr and Walter, 2006). The journals from Thomson Scientific (AH, SCI, SSCI) which are mainly English language journals, have the best coverage/identification percentage wise, at more than 80%.

## 2. Distribution of document types

We then analyzed Google Scholar data in terms of the document type to which it belongs. In total 621,000 Google Scholar records were analyzed. The Google Scholar hits can be categorized into four different types (Link, Citation, PDF-Link and other formats such as PS, DOC, RTF). The distribution of document types is closely related to the results described above. The high ratio of journals found is reflected in the high percentage of document type *Citation* (28%). This type, which Google terms "offline-record," can not be described as a classical reference because it is only comprised of extracted references and offers only minimal bibliographic information (see Figure 4). The document type *Link*; a literature reference with an abstract, appears in the analyzed data with the largest ratio at approximately 53%. The references with direct access to full-text in the pdf format (full-text) are clearly less often represented reaching only 19%. The other formats have negligible ratios. Our previous study (April/May 2005) showed similar values for both of the main document types (*Link* and *Citation*) of about 44% (compare Mayr and Walter, 2006). Based on these figures we conclude that the content coverage of the service has been expanded in 2006.

# Take in Figure 4

Fig. 4: Google Scholar results list for the query *Koelner Zeitschrift fuer Soziologie und Sozialpsychologie*

The values of the document types from the results analysis are detailed separately for each journal list in Table 2.

Table 2: Distribution of document types among the lists queried

| Lists | Link % | Citations % | Full-text % |
|---|---:|---:|---:|
| AH | 41.78 | 50.73 | 7.49 |
| DOAJ | 48.29 | 29.61 | 22.11 |
| IZ | 10.42 | 83.11 | 6.48 |
| SCI | 61.35 | 16.72 | 21.94 |
| SSCI | 49.38 | 32.84 | 17.78 |

What stands out here is that the SOLIS database journals (see IZ, German language social science journals) generate, for the most part, only citations as results (see 83.11% under document types *Citation*). The reason is that Google Scholar cannot (directly) link the mostly German language articles and so offers only the extracted references from indexed documents (see Figure 4 as an example). The ratio of citations from the international journal lists (DOAJ, AH, SCI, SSCI) is clearly lower but also, to some extent, relatively high (see lists AH with 50.7% citations). Approximately 30% of open access articles (DOAJ) could not be listed as full-text or links. The international STM journals from Thomson Scientific (SCI) display the highest percentage of link references (approximately 61%). A noticeable increase in the document type link can be seen for all lists when compared with our previous study (April/May 2005).

### 3. Distribution of web servers

If a result links to a hyperlinked reference (document type link or full-text) the distribution of this web server can be evaluated per journal list and a frequency distribution computed.

Table 3 shows the 25 servers most frequently offering journal articles of the SCI list. The description column categorizes the type of server. *Publisher* indicates a commercial server offered by an academic publisher where there is a fee for full-text downloads; *Scientific portal* stands for servers offering free references and full-texts, although they do not always link directly to the full text in every case. For some there may be more than a single appropriate description, for example, portal.acm.org is a publisher and scientific portal. *Open Access* describes open access servers which deliver full-text free of charge.

Table 3: Distribution of the 25 most frequent web servers (SCI list)

| Web server | Host Name | Description | Frequency |
|---|---|---|---|
| www.springerlink.com | Springer-Verlag | Publisher | 33,148 |
| cat.inist.fr | Catalog of the Institut de l'Information Scientifique et Technique | Scientific portal | 30,495 |
| www.ingentaconnect.com | Ingenta | Publisher | 29,273 |
| doi.wiley.com | Wiley | Publisher | 12,202 |
| www.blackwell-synergy.com | Blackwell | Publisher | 11,344 |
| www.csa.com | CSA | Publisher | 11,075 |
| www.ncbi.nlm.nih.gov | National Center for Biotechnology Information | Scientific portal | 9,404 |
| taylorandfrancis.metapress.com | Taylor & Francis Group | Publisher | 8,180 |
| linkinghub.elsevier.com | Elsevier | Publisher | 7,368 |
| adsabs.harvard.edu | Smithsonian/NASA Astrophysics Data System | Scientific portal | 4,771 |
| Links.jstor.org | JSTOR | Scientific portal | 4,279 |
| content.karger.com | Karger Publishers | Publisher | 3,500 |
| portal.acm.org | Portal of the Association for Computing Machinery | Scientific portal | 3,207 |
| ieeexplore.ieee.org | Portal IEEE | Scientific portal | 2,353 |

| www.nature.com | Nature Publishing Group | Publisher | 2,190 |
| --- | --- | --- | --- |
| link.aip.org | American Institute of Physics | Scientific portal | 2,144 |
| Pubs.acs.org | American Chemical Society | Scientific portal | 2,083 |
| www.iop.org | Institute of Physics | Scientific portal | 1,280 |
| www.liebertonline.com | Mary Ann Liebert | Publisher | 1,234 |
| www.journals.cambridge.org | Cambridge University Press | Publisher | 1,161 |
| www.journals.uchicago.edu | University of Chicago Press | Publisher | 851 |
| www.thieme-connect.com | Georg Thieme Verlag | Publisher | 689 |
| www.publish.csiro.au | CSIRO | Publisher | 672 |
| www.pubmedcentral.nih.gov | National Institute of Health | Open Access | 667 |
| pubs.rsc.org | Royal Society of Chemistry | Scientific portal | 610 |

The frequency of publishers at the top of the list which can be connected to Google Scholar's cooperation with publishers and CrossRef partners is noteworthy.

Table 4 displays the ten most frequent web servers for all queried lists (AH, DOAJ, IZ, SCI, SSCI).

Table 4: Top 10 web servers per journal list

|    | **AH** | **DOAJ** | **IZ** | **SCI** | **SSCI** |
| --- | --- | --- | --- | --- | --- |
| 1  | links.jstor.org | www.scielo.br | cat.inist.fr | www.springerlink.com | links.jstor.org |
| 2  | cat.inist.fr | cat.inist.fr | www.springerlink.com | cat.inist.fr | www.ingentaconnect.com |
| 3  | muse.jhu.edu | www.biomedcentral.com | links.jstor.org | www.ingentaconnect.com | www.springerlink.com |
| 4  | www.questia.com | www.pubmedcentral.nih.gov | cesifo.oxfordjournals.org | doi.wiley.com | cat.inist.fr |
| 5  | www.springerlink.com | www.csa.com | www.psyjournals.com | www.blackwell-synergy.com | www.eric.ed.gov |
| 6  | www.ingentaconnect.com | redalyc.uaemex.mx | www.psycontent.com | www.csa.com | taylorandfrancis.metapress.com |
| 7  | www.blackwell-synergy.com | www.bioline.org.br | www.ingentaconnect.com | www.ncbi.nlm.nih.gov | www.blackwell-synergy.com |
| 8  | taylorandfrancis.metapress.com | www.hindawi.com | www.demographic-research.org | taylorandfrancis.metapress.com | www.questia.com |
| 9  | www.eric.ed.gov | www.emis.ams.org | www.cesifo-group.de | linkinghub.elsevier.com | doi.wiley.com |
| 10 | www.journals.cambridge.org | www.scielo.cl | hsr-trans.zhsf.uni-koeln.de | adsabs.harvard.edu | ideas.repec.org |

## Conclusions

We are well aware that statements and conclusions included here possibly will need to be revised following the next Google Scholar update. All results and conclusions in this study are current and based on sample tests (100 hits per query) and are valid as of January 2007. Like the widely used, familiar search service Google Web Search, Google Scholar offers fast searching with a simple, user-friendly interface. The pros of this are that the search is free of charge and is done across interdisciplinary full-text collections. The Google Scholar approach offers some potential for literature retrieval, for example, automatic citation analysis and the ranking built up from this, and oftentimes direct downloading of full-text which is sometimes also described as a subversive feature (listing of self-archived pre- and postprints). Accurate citation analysis and webometric studies based on Google Scholar data (see e.g. Belew, 2005; Noruzi, 2005; Bar-Ilan, 2006; Kousha and Thelwall, to appear; see also Webometrics Ranking of World Universities, see http://www.webometrics.info/methodology.html) can be recommended only with some limitation due to a lot of inconsistencies and vagueness (compare Jacsó, 2006a, 2006b) in the data. Citation counts aggregated by Google Scholar may work in some fields which are covered and indexed quite well, but in other fields which are perhaps more represented by the freely accessible web these counts can be very inflated. This can mislead researchers in citation analyses basing solely on Google Scholar.

The study shows that the majority of the journals on the five lists queried can be retrieved in Google Scholar. Upon closer examination the results are relativized by the high percentage of extracted references (see Table 2, values of the document type citation). The international journals from the Thomson Scientific List (particularly from the area of STM) are fairly well covered. Analysis of the web servers shows that the majority of the analyzed hits come from publishers. It seems that preference has been given to the collections of the CrossRef partners as well as additional commercial publishers partly indexed by Google Scholar (see Table 5). As tested with the social science list (IZ) the ratio of German language journals is probably very low.

Our results show that the expanding sector of open access journals (DOAJ list) is underrepresented among the servers. Something that remains unclear is why journal articles that are freely available on web servers are not readily listed by Google Scholar even though they are searchable via the classic Google Web Search. Although Google Scholar claims to provide "scholarly articles across the web," the ratio of articles from open access journals or the full-text (eprints, preprints) is comparably low.

Concerning the question of up-to-dateness, our tests show that Google Scholar is not able to present the most current data. It appears that the index is not updated regularly. The coverage and up-to-dateness of individual, specific web servers varies greatly. Our journal list queries empirically confirm Peter Jacsó's experience (Jacsó, 2005c) concerning the coverage of Google Scholar. Although this need be qualified by stating that the service is still in beta status. However this does not entirely explain deficits such as duplicates in results data, faulty results sets and some non-scientific sources.

In comparison with many abstracting and indexing databases, Google Scholar does not offer the transparency and completeness to be expected from a scientific information resource. Google Scholar can be helpful as a supplement to retrieval in abstracting and indexing databases mainly because of its coverage of freely accessible materials.

# References

Banks, M. A. (2005), "The excitement of Google Scholar, the worry of Google Print", *Biomed Digit Libr.,* Vol. 2 No. 2, available at: http://www.bio-diglib.com/content/2/1/2

Bar-Ilan, J. (2006), "An ego-centric citation analysis of the works of Michael O. Rabin based on multiple citation indexes", *Information Processing & Management,* Vol. 42 No. 6, pp.1553-1566.

Bauer, K. and Bakkalbasi, N. (2005), "An examination of citation counts in a new scholarly communication environment", D-Lib Magazine, Vol. 11 No. 9, available at: http://www.dlib.org/dlib/september05/bauer/09bauer.html

Belew, R. K. (2005), "Scientific impact quantity and quality: Analysis of two sources of bibliographic data", available at: http://arxiv.org/abs/cs.IR/0504036

Butler, D. (2004), "Science searches shift up a gear as Google starts Scholar engine", *Nature,* Vol. 432, p 423.

Ewert, G. and Umstätter, W. (1997), Lehrbuch der Bibliotheksverwaltung, Stuttgart: Hiersemann.

Giles, J. (2005), "Science in the web age: Start your engines", *Nature,* Vol. 438 No. 7068, pp. 554-555.

Harnad, S., Brody, T., Vallières, F., Carr, L., Hitchcock. S., Gingras, Y., Oppenheim, C., Stamerjohanns, H. and Hilf, E. (2004), "The green and the gold roads to Open Access", *Nature*, available at: http://www.nature.com/nature/focus/accessdebate/21.html

Jacsó, P. (2004), "Google Scholar Beta", Thomson Gale, available at: http://www.galegroup.com/servlet/HTMLFileServlet?imprint=9999®ion=7&fileName=/reference/archive/200412/googlescholar.html

Jacsó, P. (2005a), "As we may search - Comparison of major features of the Web of Science, Scopus, and Google Scholar citation-based and citation-enhanced databases", *Current Science,* Vol. 89 No. 9, pp.1537-1547.

Jacsó, P. (2005b), "Google Scholar Beta (Redux)", Thomson Gale, available at: http://www.gale.com/servlet/HTMLFileServlet?imprint=9999®ion=7&fileName=/reference/archive/200506/google.html

Jacsó, P. (2005c), "Google Scholar: the pros and the cons", *Online Information Review,* Vol. 29 No. 2, pp. 208-214.

Jacsó, P. (2006a), "Deflated, Inflated and Phantom Citation Counts", *Online Information Review,* Vol. 30 No. 3, pp. 297-309.

Jacsó, P. (2006b), "Dubious Hit Counts and Cuckoo's Eggs", *Online Information Review,* Vol. 30 No. 2, pp.188-193.

Kousha, K. and Thelwall. M. (to appear), "Google Scholar citations and Google Web/URL citations: A multi-discipline exploratory analysis", *Journal of the American Society for Information Science and Technology*.

Lawrence, S., Giles, C. L. and Bollacker, K. (1999), "Digital Libraries and Autonomous Citation Indexing", *IEEE Computer,* Vol. 32 No. 6, pp. 67-71.

Lewandowski, D. and Mayr, P. (2006), "Exploring the academic invisible web", *Library Hi Tech*, Vol. 24 No. 4, pp. 529-539, available at: http://conference.ub.uni-bielefeld.de/2006/proceedings/lewandowski_mayr_final_web.pdf

Markoff, J. (2004), "Google Plans New Service For Scientists And Scholars", New York Times, New York.

Mayr, P. and Tosques, F. (2005), "Google Web APIs - An Instrument for Webometric Analyses?", *Proceedings of the 10th International Conference of the International Society for Scientometrics and Informetrics*, Stockholm (Sweden), available at: http://www.ib.hu-berlin.de/~mayr/arbeiten/ISSI2005_Mayr_Toques.pdf

Mayr, P. and Walter, A.-K. (2006), "Abdeckung und Aktualität des Suchdienstes Google Scholar", *Information - Wissenschaft & Praxis,* Vol. 57 No. 3, pp.133-140, available at: http://www.ib.hu-berlin.de/~mayr/arbeiten/IWP_3_06_MayrWalter.pdf

Noruzi, A. (2005), "Google Scholar: The Next Generation of Citation Indexes", *Libri,* Vol. 55, pp. 170-180.

Payne, D. (2004), "Google Scholar welcomed." *The Scientist,* Vol. 5 No. 1.

Price, G. (2004), "Google Scholar Documentation and Large PDF Files", SearchEngineWatch, available at: http://blog.searchenginewatch.com/blog/041201-105511

Sullivan, D. (2004), "Google Scholar Offers Access To Academic Information", Searchenginewatch, available at: http://searchenginewatch.com/showPage.html?page=3437471

Swan, A. and Brown, S. (2005), "Open access self-archiving: An author study", Joint Information Systems Committee (JISC), available at: http://eprints.ecs.soton.ac.uk/10999/

Terdiman, D. (2004), "A Tool for Scholars Who Like to Dig Deep", New York Times, New York.